\documentclass[prb,superscriptaddress,twocolumn]{revtex4-1}
\usepackage{times,amsmath}
\usepackage{epsfig}
\usepackage{color}
\usepackage{longtable}
\usepackage{graphicx}% Include figure files
\usepackage{dcolumn}% Align table columns on decimal point
\usepackage{bm}% bold math
\usepackage{bookmark}
\usepackage{tabularx}% bold math
\usepackage{hyperref}
\usepackage{multirow}
\hypersetup{colorlinks=true, citecolor=blue, filecolor=blue, linkcolor=blue, urlcolor=blue}

\begin{document}

\title{Size-Dependent Nucleation in Crystal Phase Transition from Machine Learning Metadynamics}

\author{Pedro A. Santos-Florez}
\affiliation{Department of Physics and Astronomy, University of Nevada, Las Vegas, NV 89154, USA}

\author{Howard Yanxon}
\affiliation{X-ray Science Division, Argonne National Laboratory, Lemont, IL 60439, USA}

\author{Byungkyun Kang}
\affiliation{Department of Physics and Astronomy, University of Nevada, Las Vegas, NV 89154, USA}

\author{Yansun Yao}
\email{yansun.yao@usask.ca}
\affiliation{Department of Physics and Engineering Physics, University of Saskatchewan, Saskatoon, Saskatchewan, Canada S7N 5E2}

\author{Qiang Zhu}
\email{qiang.zhu@unlv.edu}
\affiliation{Department of Physics and Astronomy, University of Nevada, Las Vegas, NV 89154, USA}

\date{\today}

\begin{abstract}
In this work, we present an efficient framework that combines machine learning potential (MLP) and metadynamics to explore multi-dimensional free energy surfaces for investigating solid-solid phase transition. Based on the spectral descriptors and neural networks regression, we have developed a computationally scalable MLP model to warrant an accurate interpolation of the energy surface where two phases coexist. Applying the framework to the metadynamics simulation of B4-B1 phase transition of GaN under 50 GPa with different model sizes, we observe the sequential change of phase transition mechanism from collective modes to nucleation and growths. When the system size is at or below 128 000 atoms, the nucleation and growth appear to follow a preferred direction. At larger sizes, the nucleation tends to occur at multiple sites simultaneously and grow to microstructures by passing the critical size. The observed change of atomistic mechanism manifests the importance of statistical sampling with large system size. The combination of MLP and metadynamics is likely to be applicable to a broad class of induced reconstructive phase transitions at extreme conditions.
\end{abstract}

%\pacs{71.15.Mb,74.70.Xa,74.25.Jb,71.27.+a}

\vskip 300 pt

\maketitle

%\section{INTRODUCTION}
Solid-solid phase transitions driven by pressure or temperature are the most common phase transitions in nature. These transitions are important for our understanding of matters under changing environments, such as crystal formation in geological processes. Solid-solid phase transitions are also used to manufacture new materials, i.e., steel-making, synthesis of ceramics, or creating diamond from carbon under high pressure \cite{Smith-1995}. Despite their ubiquity, however, elucidation of the microscopic mechanism of solid-solid phase transitions is significantly challenging, due to the need of \textit{in-situ} high-resolution imaging technology under extreme physical conditions. Therefore, computational studies have been devoted to investigate solid-solid phase transitions and provide insights to the microscopic mechanism. Particularly, atomistic simulations based on molecular dynamics (MD) can trace atomic motions in phase transitions, but their effectiveness is hurdled by the length and time scales allowed for these simulations\cite{Marx-2009}. Commonly used \textit{ab initio} simulations, for example, are restricted up to a few hundreds of atoms, and therefore can only mimic energy barrier crossing via concerted motion of atoms, which is naturally a rare event. Most solid-solid phase transitions are thermodynamically first order and initialized by nucleation that may proceed through intermediate states. When a nucleus of the new phase grows in a crystal, free energy is gained in the core but penalized at the interface. The competition between these two contributions results in a nucleation barrier, which the system must overcome for the nucleus to grow to a critical size (i.e., a few thousands of atoms), leading to a cascade of bulk changes \cite{Khaliullin-2011, Tian-Nature-2014}. The primary difficulty in simulating nucleation and grain growths is the requirement of a very large system size to enable statistical sampling and avoid nuclei interacting with their periodic images. A realistic material simulation must reach beyond collective atomic motions to enable such nucleation dynamics, while at the same time maintains a truthful description of the material (i.e., oppose to model simulations \cite{Qi-PRL-2015,Peng-NatMat-2014}). 

Notable progress has been made in the simulation of phase transition using metadynamics \cite{Laio-PNAS-2002,Martonak-PRL-2003} over the years. This method is designed to overcome large energy barriers through positively biased MD. In metadynamics, the free energy ($G$) is described by a number of collective variables (CVs, denoted as $\boldsymbol{s}$). In order to drive the system out of an energy well, $G$ is continuously modified by filling Gaussian potentials that discourage the revisit of already explored phase space. At timestep $t$, the total free energy is expressed as \cite{Martonak-PRL-2003} 
\begin{equation}
    G^t(\boldsymbol{s}) = G(\boldsymbol{s}) + \sum_{t'<t} W e^{-|\boldsymbol{s}-\boldsymbol{s}^{t'}|^2/2\delta s^2},
\end{equation}
where $W$ and $\delta s$ are the height and width of the Gaussian. At each metastep, the evolution of $\boldsymbol{s}$ follows
\begin{equation}
    \boldsymbol{s}^{t+1} = \boldsymbol{s}^{t} + \delta s \frac{\boldsymbol{F}^t}{|\boldsymbol{F}^t|},
\end{equation}
where $|\boldsymbol{F}^t|=-\partial G^t / \partial \boldsymbol{s}$.

Metadynamics has been used successfully to study solid-solid phase transitions in a variety of systems \cite{Martonak-ACIE-2003, Behler-PRL-2008,Yao-PRL-2009,Sun-PNAS-2009}. Recently, metadynamics incorporated with structure-factor-based CVs for long range order was successfully applied to study the crystallization process of liquids, which share similar features with solid-solid phase transitions \cite{Bonati-PRL-2018, Niu-PRL-2019}. In solid state, the simulation has been scaled up with Gaussian process regression (GAP) potential \cite{bartok2010gaussian} for a B4-B1 phase transition in GaN (up to 4096 atoms) using scaled lattice matrix as the CVs, where the onset of nucleation is revealed \cite{Tong-PRB-2021}. In a more recent work, metadyanmics simulation is carried out using classical Born-Mayer-Huggins-Fumi-Tosi (BMHFT) potentials \cite{Tosi-JPCS-1964} and two CVs, namely coordination number and volume, which achieves the simulation of B1-B2 transition in NaCl proceeding via nucleation and growth up to 64 000 atoms \cite{Matej-PRL-2021}. These studies adopt either classical or machine learning scheme to describe the free energy surface, providing a compromise between accuracy and length/time scale.

In the present study, we focus on providing a scheme for a substantially scaled up simulation of solid-solid phase transitions using machine learning representation of multi-dimensional free energy surfaces. We developed a new efficient method for training neural network (NN) potential\cite{behler-PRL-2003} on a large set of pre-computed structure-properties dataset, including structures, stress tensors and interatomic forces calculated at density functional theory (DFT) level and optimized with respect to NN parameters to achieve the best reproducibility. We show that the potential is capable of collecting all relevant information needed for solid-solid phase transitions, and the local nature of the potential is well suited to describe the nucleation dynamics. The metadynamics scheme is implemented to allow all degrees of freedom of the supercell to evolve during the phase transition, thus it enables an effective path selection in solid-solid phase transition. For the sake of clarity, the applicability of this method is demonstrated using a classic example, the B4-B1 phase transition in GaN, using a simulation box with size up to half million atoms under high pressure. The choice of the system is determined by scientific interest, since GaN as a wide band-gap semiconductor has technological importance \cite{Fasol-1997}. Multiple transition paths were proposed \cite{Limpijumnong-PRL-2001,Saitta-PRB-2004} exhibiting the complexity of its high-pressure transition paths.
The present simulation reveals the bulk phase transition through nucleation and growth, and a system size dependent crossover from directional nucleation of single cluster to homogeneous nucleation of multiple clusters. The method can be easily expanded and transferred to other systems with well-trained machine learning potentials.

%NN-SNAP and training
In the past decade, machine learning methods have been widely applied to resolve the dilemma in comprising between accuracy and cost \cite{behler2015constructing}. Machine learning potential (MLP) are trained by minimizing the cost function to attune the model to deliberately describe the \textit{ab initio} data. Among many different MLP models, both NN and GAP techniques are becoming increasingly popular in the materials modelling community. Compared to GAP, NN is more suitable for large scale simulation due to its better scalability. Very recently, we have developed the NN version of spectral neighbor analysis potential (NN-SNAP) \cite{thompson2015spectral, yanxon2020neural, zagaceta2020, yanxon2020pyxtalff} based on the bispectrum coefficient descriptors \cite{bartok2010gaussian, Bartok-PRB-2013} and implemented them to the \texttt{ML-IAP} package inside the \texttt{LAMMPS} software \cite{lammps}. To train an accurate NN-SNAP model for describing the GaN's B4-B1 transition, we start with the existing dataset from a recent work \cite{Tong-PRB-2021}. Furthermore, we apply the trained model to run NPT MD simulation for 64-atoms B1/B4 models at different pressure-temperature conditions to sample more phase space. For the representative MD configurations, we perform additional single-point DFT calculations (see Fig. S1 and Table S1 \cite{SM}) to augment the training data and improve the NN-SNAP model. Finally, we run another similar iteration based on metadynamics simulation of the 32 and 64 atoms models, ensuring that the final NN-SNAP model can also well describe the transition paths between B1 and B4 energy basins on the potential energy surface (PES).

%Validation on the small cells
\begin{figure}[ht]
\centering
\includegraphics[width=0.45 \textwidth]{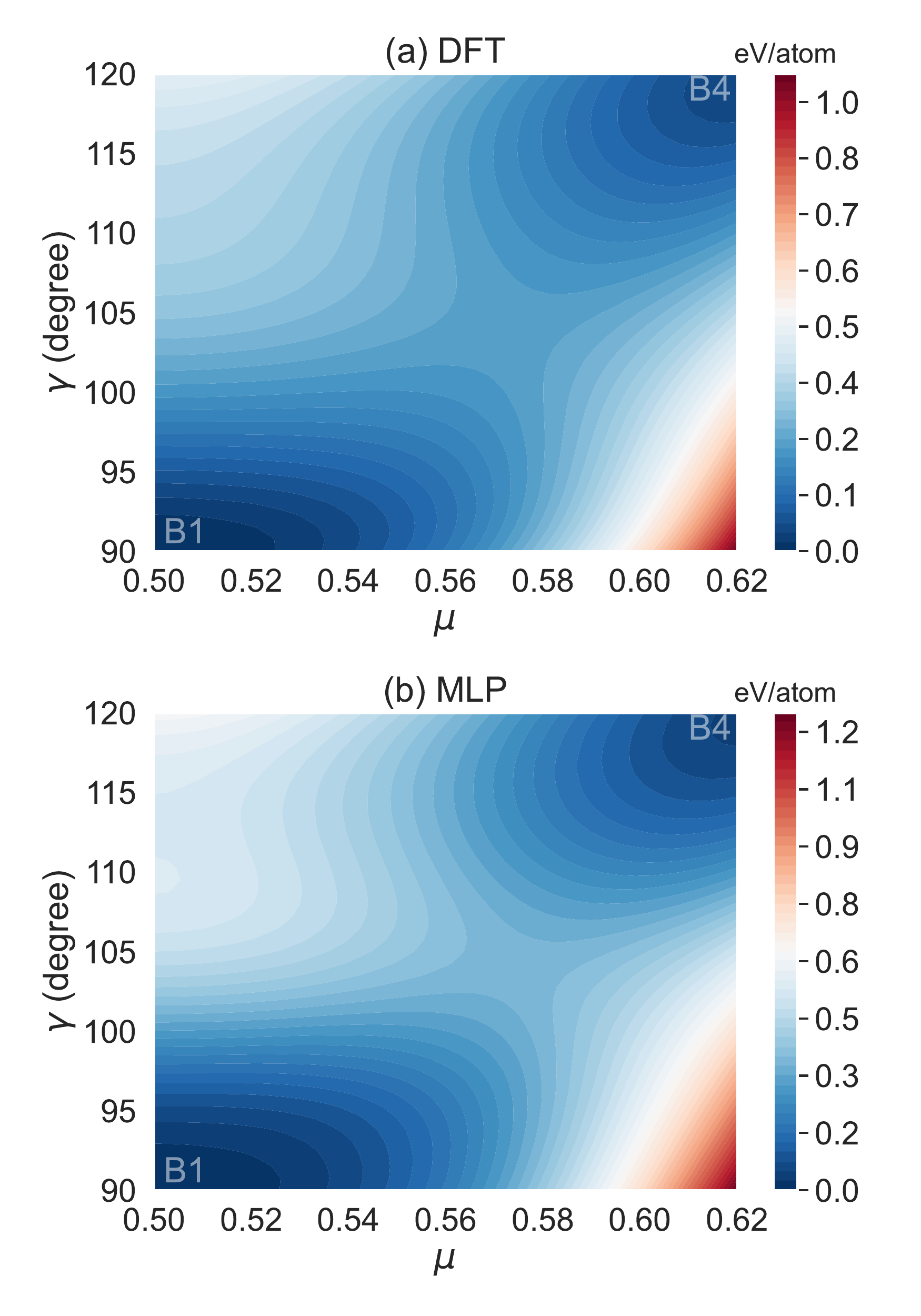}
\caption{\label{Fig1} Comparison of PES's around B1 and B4 phases of GaN at 50 GPa computed using (a) DFT and (b) MLP methods. Both PES's were computed with two order parameters, the angle $\gamma$ and sublattice spacing $\mu$, in a 4-atoms unit cell.}
\end{figure}

To check the validity of our MLP, it is necessary to examine some basic physical quantities with respect to the DFT results. In the past DFT-based studies \cite{Limpijumnong-PRL-2001, Saitta-PRB-2004, Yao-PRB-2013}, it was found that three key variables describe the B4-B1 transition of GaN, including (1) the spacing $\mu$ between Ga/N sublattices changing from 0.5 to 0.62; (2) the basal angle $\gamma$ decreasing from 120$^\circ$ to 90$^\circ$; and (3) the $c/a$ ratio reducing from 1.633 to 1.414. Multiple transition paths can be envisioned upon switching the orders of these changes, resulting in different intermediate phases. Following our earlier work \cite{Yao-PRB-2013}, we first checked the 2D PES of a 4-atoms GaN system as a function of $\mu$ and $\gamma$. As shown in Fig. \ref{Fig1}, our MLP can well reproduce the DFT's PES, despite some negligible discrepancies at the high energy regions. It appears that in both PES's the lowest energy paths are adjacent to the surface diagonal in which $\mu$ and $\gamma$ change nearly simultaneously. Since these data were not part of the dataset used in the training, the excellent agreement in blind predicting PES warrants a good interpretive capability of our MLP, which enables an accurate description of nucleation region where the two phases coexist. In addition, we checked the elastic properties and phonon dispersion for both B1 and B4 structures in a wide range of pressure and observed an overall good agreement as well (see Figs S2-S3 \cite{SM}).

%Metadynamics simulation
\begin{figure}[t]
\centering
\includegraphics[width=0.5 \textwidth]{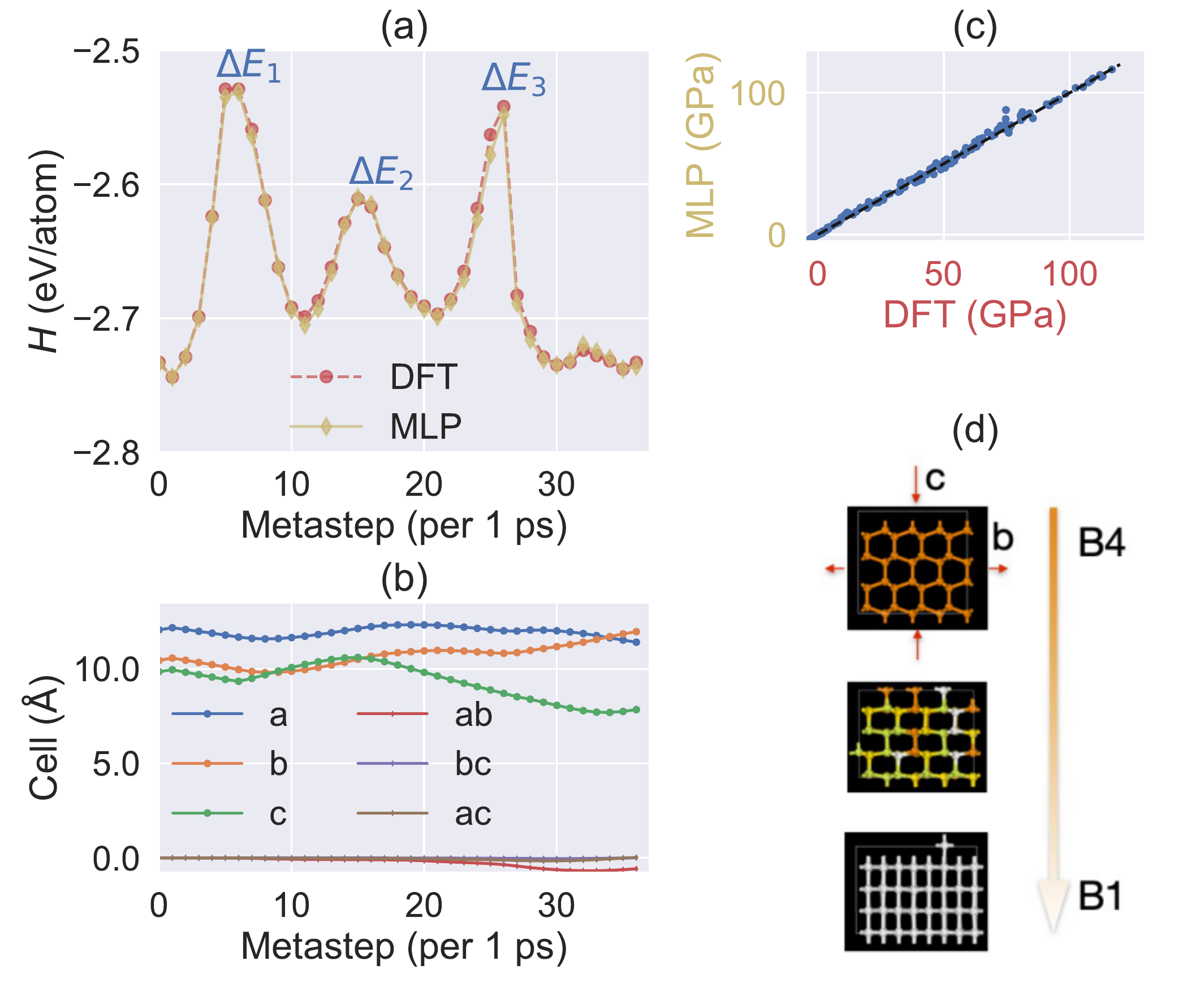}
\caption{\label{Fig2} The metadynamics simulation of GaN's B4-B1 transition for a system with 128 atoms in the supercell. (a) Evolution of enthalpy ($H$) in MLP based metadynamics (yellow) simulation. Enthalpies of selected structures along the metadynamics trajectory were also recomputed with DFT (red). (b) Corresponding evolution of CVs along the metadynamics trajectory. ($a$, $b$, $c$) and ($ab$, $bc$, $ac$) denote uniaxial and shear modes, respectively. (c) Comparison of stress tensor values from DFT and MLP calculations. (d) Outline of atomistic motions during the B4-B1 transition at critical steps. In (d) the atoms in orange follow the 4-coordinated diamond packing (B4) and atoms in white follow the 6-coordinated simple cubic packing (B1).}
\end{figure}

Encouraged by the validation results, we proceed to perform metadynamics simulations (by using the 6-dimensional cell parameters as the CVs) to explore the B4-B1 transition with large system sizes. Fig. \ref{Fig2} shows a typical trajectory (enthalpy evolution with metastep) for a system consisting of 128 atoms at 50 GPa. In this simulation, the Gaussian parameters were set as $\delta s$=0.2 \AA, and $W$=3000 GPa$\cdot$\AA$^3$, respectively. Once the evolution reached the B1 phase, we recalculated the enthalpy along the MLP trajectory with DFT. From Fig. \ref{Fig2}a, it is clear that the computed DFT energies nearly overlay with the MLP results. More importantly, the stress tensors as the driver for the change of unit cell are well reproduced in MLP (Fig. \ref{Fig2}c). The excellent agreements suggest that our MLP trained with smaller systems is equally applicable to large scale simulations at the accuracy on par with DFT.

According to Fig. \ref{Fig2}a-b, the entire simulation can be divided into three stages. Stage \textbf{i} corresponds to a simultaneous compression of $a$, $b$, $c$ axes at the first 10 metasteps. These changes clearly require a high penalty energy ($\Delta E_1$=0.204 eV/atom) and therefore the attempt along this direction quickly gives up and the unit cell bounces back. As a consequence of over-compression in \textbf{i}, stage \textbf{ii} acts in an opposite way through cell expansion. Its penalty energy is lower ($\Delta E_2$=0.122 eV/atom) and only $c$ axis starts to rebound when it reaches the 2nd peak. Afterwards, metadynamics starts to explore more diverse evolution and finds a transition path at stage \textbf{iii}, via simultaneous large compression on $c$ and small expansion on $b$ (while $a$ has little change). To complete the entire transition, the system needs to go through a sharp peak with a barrier $\Delta E_3$ of 0.191 eV/atom. At this system size, the phase transition still undergoes a concerted manner. Fig. \ref{Fig2}d presents the schematic picture following the transition mechanism within several metasteps (1 ps per metastep). The large reduction in $c$ results in a decrease of $c/a$ ratio. The increment in $b$ causes the change in basal angle $\gamma$. The internal spacing $\mu$ is progressively modified through MD equilibrium after the unit cell change in each metastep. We note that there also exists slight distortion on the inclinations shown by the minor changes of $ab$, $bc$, and $ac$. The shear modes have been found to be critical for speeding up the B4-B1 transition for small systems, where the supercell must incline to accommodate the structure change \cite{Tong-PRB-2021}. However, these shear modes become less important for large systems as we will discuss below.

Although we also found several other paths for B4-B1 transition in the metadymamics simulation, the results in Fig. \ref{Fig2}a provide a relatively simple picture as it can identify a successful path within only a few attempts, which forms a foundation for the same transition in larger systems. Using the same scaled Gaussian parameters, we then investigate the dependence of system size by varying number of atoms ($N$) from several hundreds to more than half a million.

\begin{figure}[ht]
\centering
\label{Fig3}
\includegraphics[width=0.5 \textwidth]{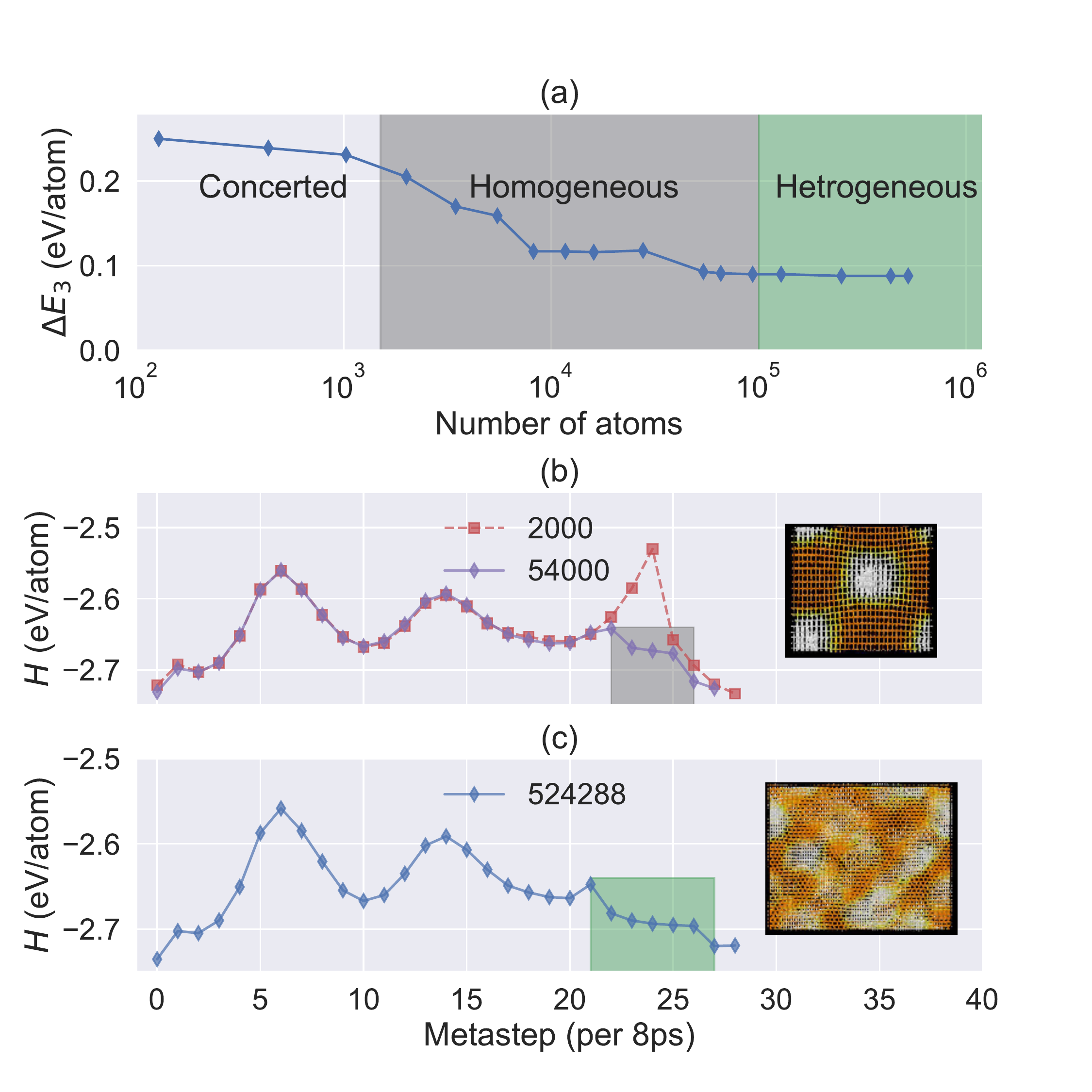}
\caption{\label{Fig3} The metadynamics simulation of GaN's B4-B1 transition for various system sizes. (a) Evolution of phase transition barrier ($\Delta E_3$) as a function of system size ($N$). (b) and (c) Time evolution of enthalpy ($H$) during the phase transition for several selected systems. Two representative inset snapshots show directional [in (b)] and homogeneous [in (c)] nucleation mechanisms. The atoms in orange follow the 4-coordinated diamond packing (B4) and atoms in white follow the 6-coordinated simple cubic packing (B1). The atoms in yellow are at the interface. Figures are generated using \texttt{OVITO}\cite{ovito}.}
\end{figure}

\begin{figure*}[ht]
\centering
\includegraphics[width=1.0 \textwidth]{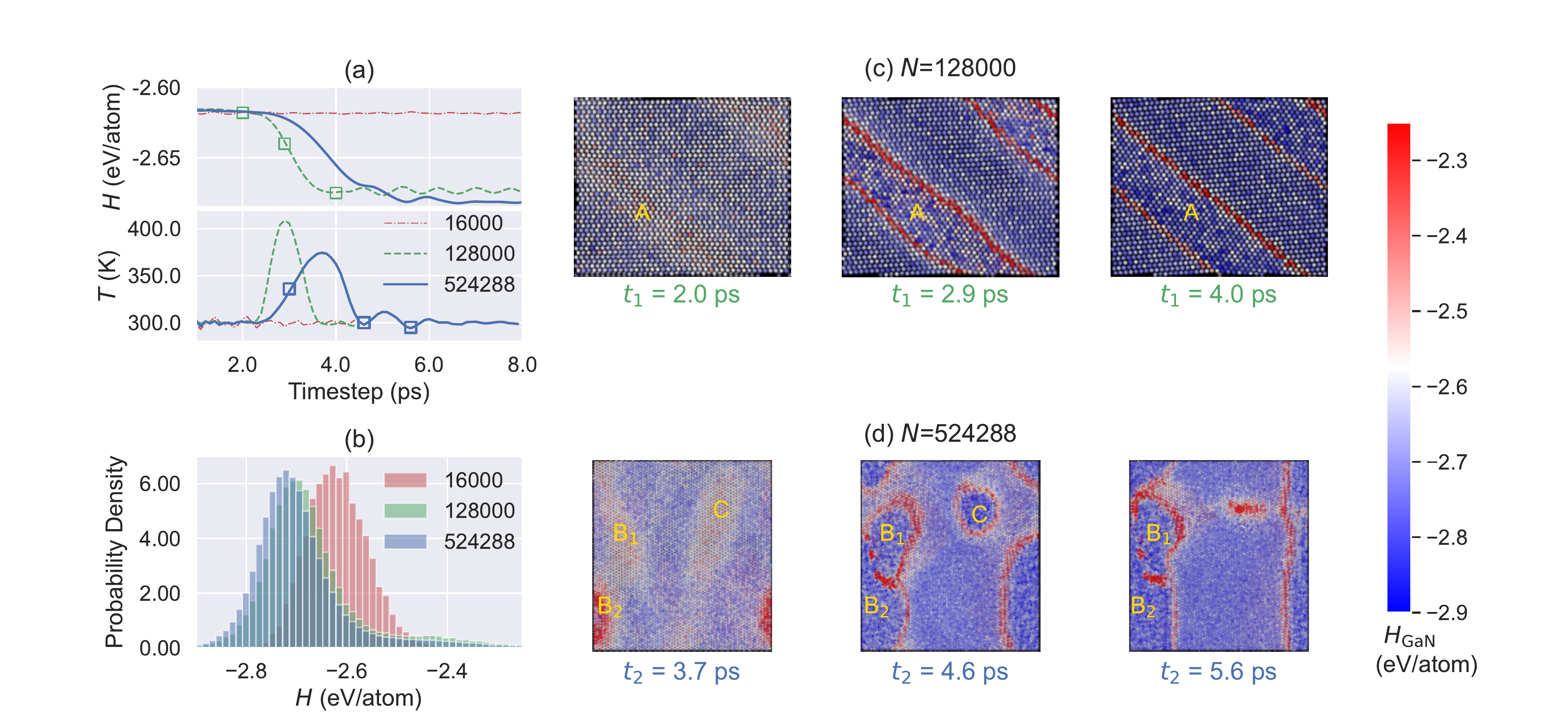}
\caption{\label{Fig4} The system size-dependent nucleation process in GaN's B4-B1 transition. (a) Variation of enthalpy and temperature as a function of time for three different systems ($N$=16 000, 128 000 and 524 288) with the same starting strain condition. (b) Distribution of atomic enthalpy ($H_\textrm{GaN}$) for equilibrated configurations. (c) and (d) Consecutive MD snapshots at selected times for $N$=128 000 and 524 288, respectively. Several representative nucleus sites (A, B$_1$, B$_2$, C) are marked in gold color to guide the eye. To compute $H_\textrm{GaN}$, the atomic volumes are estimated by Voronoi tessellation from \texttt{OVITO}\cite{ovito}.}
\end{figure*}

Table S2 \cite{SM} summarizes the details of all metadynamics simulations in this study. In general, both $\Delta E_1$ and $\Delta E_2$ quickly converge as the system goes beyond 2000 atoms (see Fig. S4). They represent two hard modes that cannot trigger the phase transition, and the converged energy corresponds to the point at which the exploration is terminated. However, $\Delta E_3$, corresponding to the barrier of low-energy transition path between B4 and B1 is found to continuously decrease with the system size. Fig. \ref{Fig3}a shows the evolution of $\Delta E_3$ as a function of system size. Up to $N$=1024, the system undergoes concerted transition in which $\Delta E_3$ steadily drops with increasing system size. This is expected since a large supercell allows more collective modes to trigger the deformation. The mechanism changes to nucleation in a 2000-atoms system (Fig. \ref{Fig3}a). From the analysis of metadynamics trajectory (Figs. S5-S7), there is a clear evidence of nucleus formation at the critical stage. The B1 nucleus for $N$=2000 grows rapidly and completes the entire transition within only one metastep (25) with an energy barrier of 0.215 eV/atom. The nucleation process is seen to be particularly effective in further reducing the barrier in larger systems. The terminal value of the energy barrier is around 0.1 eV/atom when $N$=524 288, the largest system attempted in the present study. Fig. S7 shows several representative MD snapshots for $N$=54 000. A single nucleus occurs at a smaller critical strain load. Further, the nucleus continues to grow and expand through five consecutive metasteps to complete the entire transition. Clearly, a larger simulation size allows statistically more versatile atomic fluctuations to initiate the phase transition and broaden the time window of the critical event. Therefore, localized nucleation can significantly reduce the barrier, thus providing a more realistic picture to describe the phase transition.

It is interesting to note that a similar trend was also observed in the study of B1-B2 transition of NaCl consisting of up to 64 000 atoms based on a classical potential and two CVs of cell volume and coordination number \cite{Matej-PRL-2021}. In that work, the authors claim that only a small number of CVs can be used to construct the PES and study nucleation in large systems, while the supercell-based CVs are only applicable in concerted transitions. However, our metadynamics simulations, based on the more commonly used 6-dimensional cell parameters \cite{Martonak-PRL-2003}, is able to produce the same physical picture. This may be explained by the fact that a more accurate MLP can avoid some artificial effects on the PES from the classical force fields and therefore enable a more faithful configuration exploration. Allowing all degrees of freedom of the unit cell to change, which is then coupled with atomic motions in MD, enables the simulation to statistically sample more phase space that may capture different atomic events. Therefore, we continue on to much larger systems that were not reported in the past. Up to $N$=128 000, the nucleus formation and growth is observed to always follow a preferred direction. Nevertheless, simulations with larger sizes suggests more homogeneous nucleation events. Compared to the directional nucleus picture as shown in Fig. \ref{Fig3}b, we find multiple nuclei with different sizes and orientations appearing at the early stage of phase transition (see Fig. \ref{Fig3}c).

To understand the dependence of nucleation mechanism on system size, we focus on analyzing the critical states that were found in our metadynamics simulation. We choose three representative system sizes, namely $N$=16 000, 128 000 and 524 288. From the metadynamics trajectories of these three systems, we picked critical metasteps that are around the phase transition window, and then run longer canonical ensemble NVT MD simulations up to 15 ps (as opposed to 1 ps in an ordinary metastep) with the same strain condition. This is to elongate the time window allowed for atomic nucleation in one metastep. As shown in Fig. \ref{Fig4}a, the system size clearly has a marked impact on the phase transition mechanism. While the simulation for $N$=16 000 only oscillates for the entire period, we observe the rare phase transition events for two other large systems as suggested by the sizable changes in both temperature and enthalpy. For $N$=128 000, there is a single transition window between $t_1$=2.0-4.0 ps. As shown in Fig. \ref{Fig4}c, it first undergoes sufficient atomic fluctuation to initiate a nucleus site (marked as A). The A site then continues to grow along [110] direction as shown by the expansion of interface regions (colored in red). The critical nucleus is therefore a cylindrical volume extending across the periodic boundary conditions along one dimension. Interestingly, directional growth of nucleus has been observed previously in simulation with atom numbers of this scale or below, which was attributed to the large strain close to the periodic boundary \cite{Tong-PRB-2021, Matej-PRL-2021}. This is consistent to the finding in metadynamics simulation that the drop in $\Delta E_3$ actually accompanies with a lower critical strain. While the system size seems to suppress the development of multiple nuclei, a smaller interface-to-volume ratio favors the formation of cylindrical nucleus.

%When the system increases further, the directional nucleation and growth persists till $N$=65 536, despite a slight drop in $\Delta E_3$ to 0.091 eV/atom.

On the other hand, the trajectory for $N$=524 288 exhibits a more complicated picture with three stages. First, multiple nuclei (marked as B$_1$, B$_2$ and C in Fig. \ref{Fig4}d) emerge simultaneously and grow until $t_2$= 3.0 ps, which causes the temperature of the system to rise. Further, these nuclei grow and start to interact with each other, when they get close around $t_2$=4.5--5.7 ps. These nuclei are clearly three dimensional in nature; some are not far from spheres. Closer nuclei (i.e., B$_1$ and B$_2$) merge into microstructure with fine grains and bypass the critical size. The grains reaches the largest size at peak temperature, which agrees with the classical theory of homogeneous nucleation for two competing effects with respect to temperature \cite{Okita-JCG-2017}. After this stage the number of grains decreases, and the temperature drops accordingly. Some nuclei (i.e., C) stop growing before it can reach the critical size and and eventually disappears. Similar homogeneous nucleation and microstructure formation have been observed in previous MD simulations with close to a million atoms \cite{Shibuta-SR-2015}, which manifests the importance of the statistical sampling with large system size. The competition essentially leads to the appearance of a second peak in the temperature profile and the further reduction of total energy (see Fig. \ref{Fig4}a). Finally, Fig. \ref{Fig4}b compares the distribution of atomic enthalpy ($H_\textrm{GaN}$) of the equilibrium configurations from each simulation. Clearly, a larger system size allows more versatile distribution of local strain to trigger the phase transition at different locations. Therefore, it can reduce the barrier further as compared to the single-nucleus picture.

%the Ga atoms for both configurations of $t_1$ ($N$=128000) and $t_2$ ($N$=524288) exhibit a broadened unimodal energy distribution due to thermal fluctuation. With the time evolution, the system undergoes nucleation to transform to B4 and B1 atoms, thus generating a bimodal pattern in the energy histogram. Clearly, the $t_2$-series has lower energy as compared to the $t_1$ counterparts. Therefore, it is important to note that the multiple-nucleus, that can be only found for very large system, is energetically more favorable due to statistical reason. 
%Therefore, it is important to run a l competition of different nucleation events. 
%the growth of nucleus no longer follows the cylinder shape and the transition window also becomes longer, as compared to the homogeneous nucleation mechanism. %Critical strain on $c$ axis is smaller.
%Transition window is longer (only growth of single, growth, disappearing and merge). Slope is lower.

%\subsection*{Conclusions}
In summary, we report the application of well-trained machine-learning potential to simulate solid-solid phase transitions, demonstrated in the simulation of pressure-induced B4-B1 phase transition in GaN. We trained a neural network potential based on a large set of structure-properties data and MD configurations, and validated the potential by a set of benchmark calculation based on DFT reference data. The potential is shown to adequately describe solid-solid phase transitions, particularly effective for rare events such as nucleation and growth, which paves the way to accurate atomistic modelling at a large scale. By varying the system size from several hundreds to half million atoms in metadynamics simulation, we observe sequential change of phase transition mechanism from collective modes to directional nucleation and growth, and to homogeneous nucleation at multiple sites. The combination of MLP and metadynamics is likely to be applicable to a broader class of reconstructive phase transitions at extreme conditions.

This research was sponsored by the U.S. Department of Energy, Office of Science, Office of Basic Energy Sciences, Theoretical Condensed Matter Physics program, DOE Established Program to Stimulate Competitive Research under Award Number DE-SC0021970, and Natural Sciences and Engineering Research Council of Canada (NSERC). The computing resources are provided by XSEDE (TG-DMR180040) and Compute Canada. %The code is available in \url{https://github.com/qzhu2017/***}. 
The authors thank Dr. Aidan Thompson for helpful discussions regarding the implementation of \texttt{NN-SNAP} into the \texttt{LAMMPS} package.

%\section*{SUPPLEMENTAL INFORMATION}
%Supplemental Information can be found online at
%\url{https://doi.org/10.1016/****}

%\section*{AUTHOR CONTRIBUTIONS}
%QZ proposed this idea and designed the research. 
%SB, HZ and QZ supervised this project.
%XY, KP and QZ performed and analyzed the calculations.
%All authors discussed this manuscripts.

%\section*{DECLARATION OF INTERESTS}
%The authors declare no competing financial interests.

\bibliography{ref}

\begin{thebibliography}{34}%
\makeatletter
\providecommand \@ifxundefined [1]{%
 \@ifx{#1\undefined}
}%
\providecommand \@ifnum [1]{%
 \ifnum #1\expandafter \@firstoftwo
 \else \expandafter \@secondoftwo
 \fi
}%
\providecommand \@ifx [1]{%
 \ifx #1\expandafter \@firstoftwo
 \else \expandafter \@secondoftwo
 \fi
}%
\providecommand \natexlab [1]{#1}%
\providecommand \enquote  [1]{``#1''}%
\providecommand \bibnamefont  [1]{#1}%
\providecommand \bibfnamefont [1]{#1}%
\providecommand \citenamefont [1]{#1}%
\providecommand \href@noop [0]{\@secondoftwo}%
\providecommand \href [0]{\begingroup \@sanitize@url \@href}%
\providecommand \@href[1]{\@@startlink{#1}\@@href}%
\providecommand \@@href[1]{\endgroup#1\@@endlink}%
\providecommand \@sanitize@url [0]{\catcode `\\12\catcode `\$12\catcode
  `\&12\catcode `\#12\catcode `\^12\catcode `\_12\catcode `\%12\relax}%
\providecommand \@@startlink[1]{}%
\providecommand \@@endlink[0]{}%
\providecommand \url  [0]{\begingroup\@sanitize@url \@url }%
\providecommand \@url [1]{\endgroup\@href {#1}{\urlprefix }}%
\providecommand \urlprefix  [0]{URL }%
\providecommand \Eprint [0]{\href }%
\providecommand \doibase [0]{http://dx.doi.org/}%
\providecommand \selectlanguage [0]{\@gobble}%
\providecommand \bibinfo  [0]{\@secondoftwo}%
\providecommand \bibfield  [0]{\@secondoftwo}%
\providecommand \translation [1]{[#1]}%
\providecommand \BibitemOpen [0]{}%
\providecommand \bibitemStop [0]{}%
\providecommand \bibitemNoStop [0]{.\EOS\space}%
\providecommand \EOS [0]{\spacefactor3000\relax}%
\providecommand \BibitemShut  [1]{\csname bibitem#1\endcsname}%
\let\auto@bib@innerbib\@empty
%</preamble>
\bibitem [{\citenamefont {Smith}(1996)}]{Smith-1995}%
  \BibitemOpen
  \bibfield  {author} {\bibinfo {author} {\bibfnamefont {W.~F.}\ \bibnamefont
  {Smith}},\ }\href@noop {} {\emph {\bibinfo {title} {Principles of Materials
  Science and Engineering}}}\ (\bibinfo  {publisher} {McGraw-Hill},\ \bibinfo
  {year} {1996})\BibitemShut {NoStop}%
\bibitem [{\citenamefont {Marx}\ and\ \citenamefont
  {Hutter}(2009)}]{Marx-2009}%
  \BibitemOpen
  \bibfield  {author} {\bibinfo {author} {\bibfnamefont {D.}~\bibnamefont
  {Marx}}\ and\ \bibinfo {author} {\bibfnamefont {J.}~\bibnamefont {Hutter}},\
  }\href@noop {} {\emph {\bibinfo {title} {Ab Initio Molecular Dynamics: Basic
  Theory and Advanced Methods}}}\ (\bibinfo  {publisher} {Cambridge University
  Press},\ \bibinfo {address} {Cambridge},\ \bibinfo {year} {2009})\BibitemShut
  {NoStop}%
\bibitem [{\citenamefont {Khaliullin}\ \emph {et~al.}(2011)\citenamefont
  {Khaliullin}, \citenamefont {Eshet}, \citenamefont {Kühne}, \citenamefont
  {Behler},\ and\ \citenamefont {Parrinello}}]{Khaliullin-2011}%
  \BibitemOpen
  \bibfield  {author} {\bibinfo {author} {\bibfnamefont {R.~Z.}\ \bibnamefont
  {Khaliullin}}, \bibinfo {author} {\bibfnamefont {H.}~\bibnamefont {Eshet}},
  \bibinfo {author} {\bibfnamefont {T.~D.}\ \bibnamefont {Kühne}}, \bibinfo
  {author} {\bibfnamefont {J.}~\bibnamefont {Behler}}, \ and\ \bibinfo {author}
  {\bibfnamefont {M.}~\bibnamefont {Parrinello}},\ }\href {\doibase
  10.1038/nmat3078} {\bibfield  {journal} {\bibinfo  {journal} {Nat. Mater.}\
  }\textbf {\bibinfo {volume} {10}},\ \bibinfo {pages} {693} (\bibinfo {year}
  {2011})}\BibitemShut {NoStop}%
\bibitem [{\citenamefont {Huang}\ \emph {et~al.}(2014)\citenamefont {Huang},
  \citenamefont {Yu}, \citenamefont {Xu}, \citenamefont {Hu}, \citenamefont
  {Ma}, \citenamefont {Wang}, \citenamefont {Zhao}, \citenamefont {Wen},
  \citenamefont {He}, \citenamefont {Z.},\ and\ \citenamefont
  {Y.}}]{Tian-Nature-2014}%
  \BibitemOpen
  \bibfield  {author} {\bibinfo {author} {\bibfnamefont {Q.}~\bibnamefont
  {Huang}}, \bibinfo {author} {\bibfnamefont {D.}~\bibnamefont {Yu}}, \bibinfo
  {author} {\bibfnamefont {B.}~\bibnamefont {Xu}}, \bibinfo {author}
  {\bibfnamefont {W.}~\bibnamefont {Hu}}, \bibinfo {author} {\bibfnamefont
  {Y.}~\bibnamefont {Ma}}, \bibinfo {author} {\bibfnamefont {Y.}~\bibnamefont
  {Wang}}, \bibinfo {author} {\bibfnamefont {Z.}~\bibnamefont {Zhao}}, \bibinfo
  {author} {\bibfnamefont {B.}~\bibnamefont {Wen}}, \bibinfo {author}
  {\bibfnamefont {J.}~\bibnamefont {He}}, \bibinfo {author} {\bibfnamefont
  {L.}~\bibnamefont {Z.}}, \ and\ \bibinfo {author} {\bibfnamefont
  {T.}~\bibnamefont {Y.}},\ }\href {\doibase 10.1038/nature13381} {\bibfield
  {journal} {\bibinfo  {journal} {Nature}\ }\textbf {\bibinfo {volume} {510}},\
  \bibinfo {pages} {250} (\bibinfo {year} {2014})}\BibitemShut {NoStop}%
\bibitem [{\citenamefont {Qi}\ \emph {et~al.}(2015)\citenamefont {Qi},
  \citenamefont {Peng}, \citenamefont {Han}, \citenamefont {Bowles},\ and\
  \citenamefont {Dijkstra}}]{Qi-PRL-2015}%
  \BibitemOpen
  \bibfield  {author} {\bibinfo {author} {\bibfnamefont {W.}~\bibnamefont
  {Qi}}, \bibinfo {author} {\bibfnamefont {Y.}~\bibnamefont {Peng}}, \bibinfo
  {author} {\bibfnamefont {Y.}~\bibnamefont {Han}}, \bibinfo {author}
  {\bibfnamefont {R.~K.}\ \bibnamefont {Bowles}}, \ and\ \bibinfo {author}
  {\bibfnamefont {M.}~\bibnamefont {Dijkstra}},\ }\href {\doibase
  10.1103/PhysRevLett.115.185701} {\bibfield  {journal} {\bibinfo  {journal}
  {Phys. Rev. Lett.}\ }\textbf {\bibinfo {volume} {115}},\ \bibinfo {pages}
  {185701} (\bibinfo {year} {2015})}\BibitemShut {NoStop}%
\bibitem [{\citenamefont {Peng}\ \emph {et~al.}(2014)\citenamefont {Peng},
  \citenamefont {Wang}, \citenamefont {Wang}, \citenamefont {Alsayed},
  \citenamefont {Zhang}, \citenamefont {Yodh},\ and\ \citenamefont
  {Han}}]{Peng-NatMat-2014}%
  \BibitemOpen
  \bibfield  {author} {\bibinfo {author} {\bibfnamefont {Y.}~\bibnamefont
  {Peng}}, \bibinfo {author} {\bibfnamefont {F.}~\bibnamefont {Wang}}, \bibinfo
  {author} {\bibfnamefont {Z.}~\bibnamefont {Wang}}, \bibinfo {author}
  {\bibfnamefont {A.~M.}\ \bibnamefont {Alsayed}}, \bibinfo {author}
  {\bibfnamefont {Z.}~\bibnamefont {Zhang}}, \bibinfo {author} {\bibfnamefont
  {A.~G.}\ \bibnamefont {Yodh}}, \ and\ \bibinfo {author} {\bibfnamefont
  {Y.}~\bibnamefont {Han}},\ }\href {\doibase 10.1038/nmat4083} {\bibfield
  {journal} {\bibinfo  {journal} {Nat. Mater.}\ }\textbf {\bibinfo {volume}
  {14}},\ \bibinfo {pages} {101} (\bibinfo {year} {2014})}\BibitemShut
  {NoStop}%
\bibitem [{\citenamefont {Laio}\ and\ \citenamefont
  {Parrinello}(2002)}]{Laio-PNAS-2002}%
  \BibitemOpen
  \bibfield  {author} {\bibinfo {author} {\bibfnamefont {A.}~\bibnamefont
  {Laio}}\ and\ \bibinfo {author} {\bibfnamefont {M.}~\bibnamefont
  {Parrinello}},\ }\href {\doibase 10.1073/pnas.202427399} {\bibfield
  {journal} {\bibinfo  {journal} {Proc. Natl Acad. Sci. USA}\ }\textbf
  {\bibinfo {volume} {99}},\ \bibinfo {pages} {12562} (\bibinfo {year}
  {2002})}\BibitemShut {NoStop}%
\bibitem [{\citenamefont {Marto{\v{n}}{\'a}k}\ \emph
  {et~al.}(2003)\citenamefont {Marto{\v{n}}{\'a}k}, \citenamefont {Laio},\ and\
  \citenamefont {Parrinello}}]{Martonak-PRL-2003}%
  \BibitemOpen
  \bibfield  {author} {\bibinfo {author} {\bibfnamefont {R.}~\bibnamefont
  {Marto{\v{n}}{\'a}k}}, \bibinfo {author} {\bibfnamefont {A.}~\bibnamefont
  {Laio}}, \ and\ \bibinfo {author} {\bibfnamefont {M.}~\bibnamefont
  {Parrinello}},\ }\href {\doibase 10.1103/PhysRevLett.90.075503} {\bibfield
  {journal} {\bibinfo  {journal} {Phys. Rev. Lett.}\ }\textbf {\bibinfo
  {volume} {90}},\ \bibinfo {pages} {075503} (\bibinfo {year}
  {2003})}\BibitemShut {NoStop}%
\bibitem [{\citenamefont {Raiteri}\ \emph {et~al.}(2005)\citenamefont
  {Raiteri}, \citenamefont {Marto{\v{n}}{\'a}k},\ and\ \citenamefont
  {Parrinello}}]{Martonak-ACIE-2003}%
  \BibitemOpen
  \bibfield  {author} {\bibinfo {author} {\bibfnamefont {P.}~\bibnamefont
  {Raiteri}}, \bibinfo {author} {\bibfnamefont {R.}~\bibnamefont
  {Marto{\v{n}}{\'a}k}}, \ and\ \bibinfo {author} {\bibfnamefont
  {M.}~\bibnamefont {Parrinello}},\ }\href {\doibase 10.1002/anie.200462760}
  {\bibfield  {journal} {\bibinfo  {journal} {Angew. Chem. Int. Ed.}\ }\textbf
  {\bibinfo {volume} {44}},\ \bibinfo {pages} {3769} (\bibinfo {year}
  {2005})}\BibitemShut {NoStop}%
\bibitem [{\citenamefont {Behler}\ \emph {et~al.}(2008)\citenamefont {Behler},
  \citenamefont {Marto{\v{n}}{\'a}k}, \citenamefont {Donadio},\ and\
  \citenamefont {Parrinello}}]{Behler-PRL-2008}%
  \BibitemOpen
  \bibfield  {author} {\bibinfo {author} {\bibfnamefont {J.}~\bibnamefont
  {Behler}}, \bibinfo {author} {\bibfnamefont {R.}~\bibnamefont
  {Marto{\v{n}}{\'a}k}}, \bibinfo {author} {\bibfnamefont {D.}~\bibnamefont
  {Donadio}}, \ and\ \bibinfo {author} {\bibfnamefont {M.}~\bibnamefont
  {Parrinello}},\ }\href {\doibase 10.1103/PhysRevLett.100.185501} {\bibfield
  {journal} {\bibinfo  {journal} {Phys. Rev. Lett.}\ }\textbf {\bibinfo
  {volume} {100}},\ \bibinfo {pages} {185501} (\bibinfo {year}
  {2008})}\BibitemShut {NoStop}%
\bibitem [{\citenamefont {Yao}\ \emph {et~al.}(2009)\citenamefont {Yao},
  \citenamefont {Klug}, \citenamefont {Sun},\ and\ \citenamefont
  {Marto{\v{n}}{\'a}k}}]{Yao-PRL-2009}%
  \BibitemOpen
  \bibfield  {author} {\bibinfo {author} {\bibfnamefont {Y.}~\bibnamefont
  {Yao}}, \bibinfo {author} {\bibfnamefont {D.~D.}\ \bibnamefont {Klug}},
  \bibinfo {author} {\bibfnamefont {J.}~\bibnamefont {Sun}}, \ and\ \bibinfo
  {author} {\bibfnamefont {R.}~\bibnamefont {Marto{\v{n}}{\'a}k}},\ }\href
  {\doibase 10.1103/PhysRevLett.103.055503} {\bibfield  {journal} {\bibinfo
  {journal} {Phys. Rev. Lett.}\ }\textbf {\bibinfo {volume} {103}},\ \bibinfo
  {pages} {055503} (\bibinfo {year} {2009})}\BibitemShut {NoStop}%
\bibitem [{\citenamefont {Sun}\ \emph {et~al.}(2009)\citenamefont {Sun},
  \citenamefont {Klug}, \citenamefont {Marto{\v{n}}{\'a}k}, \citenamefont
  {Montoya}, \citenamefont {Lee}, \citenamefont {Scandolo},\ and\ \citenamefont
  {Tosatti}}]{Sun-PNAS-2009}%
  \BibitemOpen
  \bibfield  {author} {\bibinfo {author} {\bibfnamefont {J.}~\bibnamefont
  {Sun}}, \bibinfo {author} {\bibfnamefont {D.~D.}\ \bibnamefont {Klug}},
  \bibinfo {author} {\bibfnamefont {R.}~\bibnamefont {Marto{\v{n}}{\'a}k}},
  \bibinfo {author} {\bibfnamefont {J.~A.}\ \bibnamefont {Montoya}}, \bibinfo
  {author} {\bibfnamefont {M.-S.}\ \bibnamefont {Lee}}, \bibinfo {author}
  {\bibfnamefont {S.}~\bibnamefont {Scandolo}}, \ and\ \bibinfo {author}
  {\bibfnamefont {E.}~\bibnamefont {Tosatti}},\ }\href {\doibase
  10.1073/pnas.0812624106} {\bibfield  {journal} {\bibinfo  {journal} {Proc.
  Natl. Acad. Sci.}\ }\textbf {\bibinfo {volume} {106}},\ \bibinfo {pages}
  {6077} (\bibinfo {year} {2009})}\BibitemShut {NoStop}%
\bibitem [{\citenamefont {Bonati}\ and\ \citenamefont
  {Parrinello}(2018)}]{Bonati-PRL-2018}%
  \BibitemOpen
  \bibfield  {author} {\bibinfo {author} {\bibfnamefont {L.}~\bibnamefont
  {Bonati}}\ and\ \bibinfo {author} {\bibfnamefont {M.}~\bibnamefont
  {Parrinello}},\ }\href {\doibase 10.1103/PhysRevLett.121.265701} {\bibfield
  {journal} {\bibinfo  {journal} {Phys. Rev. Lett.}\ }\textbf {\bibinfo
  {volume} {121}},\ \bibinfo {pages} {265701} (\bibinfo {year}
  {2018})}\BibitemShut {NoStop}%
\bibitem [{\citenamefont {Niu}\ \emph {et~al.}(2019)\citenamefont {Niu},
  \citenamefont {Yang},\ and\ \citenamefont {Parrinello}}]{Niu-PRL-2019}%
  \BibitemOpen
  \bibfield  {author} {\bibinfo {author} {\bibfnamefont {H.}~\bibnamefont
  {Niu}}, \bibinfo {author} {\bibfnamefont {Y.~I.}\ \bibnamefont {Yang}}, \
  and\ \bibinfo {author} {\bibfnamefont {M.}~\bibnamefont {Parrinello}},\
  }\href {\doibase 10.1103/PhysRevLett.122.245501} {\bibfield  {journal}
  {\bibinfo  {journal} {Phys. Rev. Lett.}\ }\textbf {\bibinfo {volume} {122}},\
  \bibinfo {pages} {245501} (\bibinfo {year} {2019})}\BibitemShut {NoStop}%
\bibitem [{\citenamefont {Bart{\'o}k}\ \emph {et~al.}(2010)\citenamefont
  {Bart{\'o}k}, \citenamefont {Payne}, \citenamefont {Kondor},\ and\
  \citenamefont {Cs{\'a}nyi}}]{bartok2010gaussian}%
  \BibitemOpen
  \bibfield  {author} {\bibinfo {author} {\bibfnamefont {A.~P.}\ \bibnamefont
  {Bart{\'o}k}}, \bibinfo {author} {\bibfnamefont {M.~C.}\ \bibnamefont
  {Payne}}, \bibinfo {author} {\bibfnamefont {R.}~\bibnamefont {Kondor}}, \
  and\ \bibinfo {author} {\bibfnamefont {G.}~\bibnamefont {Cs{\'a}nyi}},\
  }\href {\doibase 10.1103/PhysRevLett.104.136403} {\bibfield  {journal}
  {\bibinfo  {journal} {Phys. Rev. Lett.}\ }\textbf {\bibinfo {volume} {104}},\
  \bibinfo {pages} {136403} (\bibinfo {year} {2010})}\BibitemShut {NoStop}%
\bibitem [{\citenamefont {Tong}\ \emph {et~al.}(2021)\citenamefont {Tong},
  \citenamefont {Luo}, \citenamefont {Adeleke}, \citenamefont {Gao},
  \citenamefont {Xie}, \citenamefont {Liu}, \citenamefont {Li}, \citenamefont
  {Wang}, \citenamefont {Lv}, \citenamefont {Yao},\ and\ \citenamefont
  {Ma}}]{Tong-PRB-2021}%
  \BibitemOpen
  \bibfield  {author} {\bibinfo {author} {\bibfnamefont {Q.}~\bibnamefont
  {Tong}}, \bibinfo {author} {\bibfnamefont {X.}~\bibnamefont {Luo}}, \bibinfo
  {author} {\bibfnamefont {A.~A.}\ \bibnamefont {Adeleke}}, \bibinfo {author}
  {\bibfnamefont {P.}~\bibnamefont {Gao}}, \bibinfo {author} {\bibfnamefont
  {Y.}~\bibnamefont {Xie}}, \bibinfo {author} {\bibfnamefont {H.}~\bibnamefont
  {Liu}}, \bibinfo {author} {\bibfnamefont {Q.}~\bibnamefont {Li}}, \bibinfo
  {author} {\bibfnamefont {Y.}~\bibnamefont {Wang}}, \bibinfo {author}
  {\bibfnamefont {J.}~\bibnamefont {Lv}}, \bibinfo {author} {\bibfnamefont
  {Y.}~\bibnamefont {Yao}}, \ and\ \bibinfo {author} {\bibfnamefont
  {Y.}~\bibnamefont {Ma}},\ }\href {\doibase 10.1103/PhysRevB.103.054107}
  {\bibfield  {journal} {\bibinfo  {journal} {Phys. Rev. B}\ }\textbf {\bibinfo
  {volume} {103}},\ \bibinfo {pages} {054107} (\bibinfo {year}
  {2021})}\BibitemShut {NoStop}%
\bibitem [{\citenamefont {Tosi}\ and\ \citenamefont
  {Fumi}(1964)}]{Tosi-JPCS-1964}%
  \BibitemOpen
  \bibfield  {author} {\bibinfo {author} {\bibfnamefont {M.~P.}\ \bibnamefont
  {Tosi}}\ and\ \bibinfo {author} {\bibfnamefont {F.~G.}\ \bibnamefont
  {Fumi}},\ }\href {\doibase 10.1016/0022-3697(64)90160-X} {\bibfield
  {journal} {\bibinfo  {journal} {J. Phys. Chem. Solids}\ }\textbf {\bibinfo
  {volume} {25}},\ \bibinfo {pages} {45} (\bibinfo {year} {1964})}\BibitemShut
  {NoStop}%
\bibitem [{\citenamefont {Badin}\ and\ \citenamefont
  {Marto\ifmmode~\check{n}\else \v{n}\fi{}\'ak}(2021)}]{Matej-PRL-2021}%
  \BibitemOpen
  \bibfield  {author} {\bibinfo {author} {\bibfnamefont {M.}~\bibnamefont
  {Badin}}\ and\ \bibinfo {author} {\bibfnamefont {R.}~\bibnamefont
  {Marto\ifmmode~\check{n}\else \v{n}\fi{}\'ak}},\ }\href {\doibase
  10.1103/PhysRevLett.127.105701} {\bibfield  {journal} {\bibinfo  {journal}
  {Phys. Rev. Lett.}\ }\textbf {\bibinfo {volume} {127}},\ \bibinfo {pages}
  {105701} (\bibinfo {year} {2021})}\BibitemShut {NoStop}%
\bibitem [{\citenamefont {Behler}\ and\ \citenamefont
  {Parrinello}(2007)}]{behler-PRL-2003}%
  \BibitemOpen
  \bibfield  {author} {\bibinfo {author} {\bibfnamefont {J.}~\bibnamefont
  {Behler}}\ and\ \bibinfo {author} {\bibfnamefont {M.}~\bibnamefont
  {Parrinello}},\ }\href {\doibase 10.1103/PhysRevLett.98.146401} {\bibfield
  {journal} {\bibinfo  {journal} {Phys. Rev. Lett.}\ }\textbf {\bibinfo
  {volume} {98}},\ \bibinfo {pages} {146401} (\bibinfo {year}
  {2007})}\BibitemShut {NoStop}%
\bibitem [{\citenamefont {Nakamura}\ and\ \citenamefont
  {Fasol}(1997)}]{Fasol-1997}%
  \BibitemOpen
  \bibfield  {author} {\bibinfo {author} {\bibfnamefont {S.}~\bibnamefont
  {Nakamura}}\ and\ \bibinfo {author} {\bibfnamefont {G.}~\bibnamefont
  {Fasol}},\ }\href@noop {} {\emph {\bibinfo {title} {The Blue Laser Diode}}}\
  (\bibinfo  {publisher} {Springer Verlag},\ \bibinfo {address} {Berlin},\
  \bibinfo {year} {1997})\BibitemShut {NoStop}%
\bibitem [{\citenamefont {Limpijumnong}\ and\ \citenamefont
  {Lambrecht}(2001)}]{Limpijumnong-PRL-2001}%
  \BibitemOpen
  \bibfield  {author} {\bibinfo {author} {\bibfnamefont {S.}~\bibnamefont
  {Limpijumnong}}\ and\ \bibinfo {author} {\bibfnamefont {W.~R.~L.}\
  \bibnamefont {Lambrecht}},\ }\href {\doibase 10.1103/PhysRevLett.86.91}
  {\bibfield  {journal} {\bibinfo  {journal} {Phys. Rev. Lett.}\ }\textbf
  {\bibinfo {volume} {86}},\ \bibinfo {pages} {91} (\bibinfo {year}
  {2001})}\BibitemShut {NoStop}%
\bibitem [{\citenamefont {Saitta}\ and\ \citenamefont
  {Decremps}(2004)}]{Saitta-PRB-2004}%
  \BibitemOpen
  \bibfield  {author} {\bibinfo {author} {\bibfnamefont {A.~M.}\ \bibnamefont
  {Saitta}}\ and\ \bibinfo {author} {\bibfnamefont {F.}~\bibnamefont
  {Decremps}},\ }\href {\doibase 10.1103/PhysRevB.70.035214} {\bibfield
  {journal} {\bibinfo  {journal} {Phys. Rev. B}\ }\textbf {\bibinfo {volume}
  {70}},\ \bibinfo {pages} {035214} (\bibinfo {year} {2004})}\BibitemShut
  {NoStop}%
\bibitem [{\citenamefont {Behler}(2015)}]{behler2015constructing}%
  \BibitemOpen
  \bibfield  {author} {\bibinfo {author} {\bibfnamefont {J.}~\bibnamefont
  {Behler}},\ }\href {\doibase https://doi.org/10.1002/qua.24890} {\bibfield
  {journal} {\bibinfo  {journal} {Int. J. Quantum Chem.}\ }\textbf {\bibinfo
  {volume} {115}},\ \bibinfo {pages} {1032} (\bibinfo {year}
  {2015})}\BibitemShut {NoStop}%
\bibitem [{\citenamefont {Thompson}\ \emph {et~al.}(2015)\citenamefont
  {Thompson}, \citenamefont {Swiler}, \citenamefont {Trott}, \citenamefont
  {Foiles},\ and\ \citenamefont {Tucker}}]{thompson2015spectral}%
  \BibitemOpen
  \bibfield  {author} {\bibinfo {author} {\bibfnamefont {A.~P.}\ \bibnamefont
  {Thompson}}, \bibinfo {author} {\bibfnamefont {L.~P.}\ \bibnamefont
  {Swiler}}, \bibinfo {author} {\bibfnamefont {C.~R.}\ \bibnamefont {Trott}},
  \bibinfo {author} {\bibfnamefont {S.~M.}\ \bibnamefont {Foiles}}, \ and\
  \bibinfo {author} {\bibfnamefont {G.~J.}\ \bibnamefont {Tucker}},\ }\href
  {\doibase 10.1016/j.jcp.2014.12.018} {\bibfield  {journal} {\bibinfo
  {journal} {J. Comput. Phys.}\ }\textbf {\bibinfo {volume} {285}},\ \bibinfo
  {pages} {316} (\bibinfo {year} {2015})}\BibitemShut {NoStop}%
\bibitem [{\citenamefont {Yanxon}\ \emph
  {et~al.}(2020{\natexlab{a}})\citenamefont {Yanxon}, \citenamefont {Zagaceta},
  \citenamefont {Wood},\ and\ \citenamefont {Zhu}}]{yanxon2020neural}%
  \BibitemOpen
  \bibfield  {author} {\bibinfo {author} {\bibfnamefont {H.}~\bibnamefont
  {Yanxon}}, \bibinfo {author} {\bibfnamefont {D.}~\bibnamefont {Zagaceta}},
  \bibinfo {author} {\bibfnamefont {B.~C.}\ \bibnamefont {Wood}}, \ and\
  \bibinfo {author} {\bibfnamefont {Q.}~\bibnamefont {Zhu}},\ }\href {\doibase
  10.1063/5.0014677} {\bibfield  {journal} {\bibinfo  {journal} {J. Chem.
  Phys.}\ }\textbf {\bibinfo {volume} {153}},\ \bibinfo {pages} {054118}
  (\bibinfo {year} {2020}{\natexlab{a}})}\BibitemShut {NoStop}%
\bibitem [{\citenamefont {Zagaceta}\ \emph {et~al.}(2020)\citenamefont
  {Zagaceta}, \citenamefont {Yanxon},\ and\ \citenamefont
  {Zhu}}]{zagaceta2020}%
  \BibitemOpen
  \bibfield  {author} {\bibinfo {author} {\bibfnamefont {D.}~\bibnamefont
  {Zagaceta}}, \bibinfo {author} {\bibfnamefont {H.}~\bibnamefont {Yanxon}}, \
  and\ \bibinfo {author} {\bibfnamefont {Q.}~\bibnamefont {Zhu}},\ }\href
  {\doibase 10.1063/5.0013208} {\bibfield  {journal} {\bibinfo  {journal} {J.
  Appl. Phys.}\ }\textbf {\bibinfo {volume} {128}},\ \bibinfo {pages} {045113}
  (\bibinfo {year} {2020})}\BibitemShut {NoStop}%
\bibitem [{\citenamefont {Yanxon}\ \emph
  {et~al.}(2020{\natexlab{b}})\citenamefont {Yanxon}, \citenamefont {Zagaceta},
  \citenamefont {Tang}, \citenamefont {Matteson},\ and\ \citenamefont
  {Zhu}}]{yanxon2020pyxtalff}%
  \BibitemOpen
  \bibfield  {author} {\bibinfo {author} {\bibfnamefont {H.}~\bibnamefont
  {Yanxon}}, \bibinfo {author} {\bibfnamefont {D.}~\bibnamefont {Zagaceta}},
  \bibinfo {author} {\bibfnamefont {B.}~\bibnamefont {Tang}}, \bibinfo {author}
  {\bibfnamefont {D.~S.}\ \bibnamefont {Matteson}}, \ and\ \bibinfo {author}
  {\bibfnamefont {Q.}~\bibnamefont {Zhu}},\ }\href {\doibase
  10.1088/2632-2153/abc940} {\bibfield  {journal} {\bibinfo  {journal} {Machine
  Learning: Sci. Tech.}\ } (\bibinfo {year} {2020}{\natexlab{b}}),\
  10.1088/2632-2153/abc940}\BibitemShut {NoStop}%
\bibitem [{\citenamefont {Bart\'ok}\ \emph {et~al.}(2013)\citenamefont
  {Bart\'ok}, \citenamefont {Kondor},\ and\ \citenamefont
  {Cs\'anyi}}]{Bartok-PRB-2013}%
  \BibitemOpen
  \bibfield  {author} {\bibinfo {author} {\bibfnamefont {A.~P.}\ \bibnamefont
  {Bart\'ok}}, \bibinfo {author} {\bibfnamefont {R.}~\bibnamefont {Kondor}}, \
  and\ \bibinfo {author} {\bibfnamefont {G.}~\bibnamefont {Cs\'anyi}},\ }\href
  {\doibase 10.1103/PhysRevB.87.184115} {\bibfield  {journal} {\bibinfo
  {journal} {Phys. Rev. B}\ }\textbf {\bibinfo {volume} {87}},\ \bibinfo
  {pages} {184115} (\bibinfo {year} {2013})}\BibitemShut {NoStop}%
\bibitem [{\citenamefont {Plimpton}(1995)}]{lammps}%
  \BibitemOpen
  \bibfield  {author} {\bibinfo {author} {\bibfnamefont {S.}~\bibnamefont
  {Plimpton}},\ }\href {\doibase 10.1006/jcph.1995.1039} {\bibfield  {journal}
  {\bibinfo  {journal} {J. Comput. Phys.}\ }\textbf {\bibinfo {volume} {117}},\
  \bibinfo {pages} {1} (\bibinfo {year} {1995})}\BibitemShut {NoStop}%
\bibitem [{SM()}]{SM}%
  \BibitemOpen
  \href@noop {} {}\bibinfo {note} {See Supplemental Material at
  \url{http://link.aps.org/****} for a detailed description of MLP training,
  Metadynamics simulation, energy analysis and other computational
  details}\BibitemShut {NoStop}%
\bibitem [{\citenamefont {Yao}\ and\ \citenamefont
  {Klug}(2013)}]{Yao-PRB-2013}%
  \BibitemOpen
  \bibfield  {author} {\bibinfo {author} {\bibfnamefont {Y.}~\bibnamefont
  {Yao}}\ and\ \bibinfo {author} {\bibfnamefont {D.~D.}\ \bibnamefont {Klug}},\
  }\href {\doibase 10.1103/PhysRevB.88.014113} {\bibfield  {journal} {\bibinfo
  {journal} {Phys. Rev. B}\ }\textbf {\bibinfo {volume} {88}},\ \bibinfo
  {pages} {014113} (\bibinfo {year} {2013})}\BibitemShut {NoStop}%
\bibitem [{\citenamefont {Stukowski}(2009)}]{ovito}%
  \BibitemOpen
  \bibfield  {author} {\bibinfo {author} {\bibfnamefont {A.}~\bibnamefont
  {Stukowski}},\ }\href {\doibase 10.1088/0965-0393/18/1/015012} {\bibfield
  {journal} {\bibinfo  {journal} {Model. Simul. Mater. Sic. Eng.}\ }\textbf
  {\bibinfo {volume} {18}},\ \bibinfo {pages} {015012} (\bibinfo {year}
  {2009})}\BibitemShut {NoStop}%
\bibitem [{\citenamefont {Okita}\ \emph {et~al.}(2017)\citenamefont {Okita},
  \citenamefont {Verestek}, \citenamefont {Sakane}, \citenamefont {Takaki},
  \citenamefont {Ohno},\ and\ \citenamefont {Shibuta}}]{Okita-JCG-2017}%
  \BibitemOpen
  \bibfield  {author} {\bibinfo {author} {\bibfnamefont {S.}~\bibnamefont
  {Okita}}, \bibinfo {author} {\bibfnamefont {W.}~\bibnamefont {Verestek}},
  \bibinfo {author} {\bibfnamefont {S.}~\bibnamefont {Sakane}}, \bibinfo
  {author} {\bibfnamefont {T.}~\bibnamefont {Takaki}}, \bibinfo {author}
  {\bibfnamefont {M.}~\bibnamefont {Ohno}}, \ and\ \bibinfo {author}
  {\bibfnamefont {Y.}~\bibnamefont {Shibuta}},\ }\href {\doibase
  10.1016/j.jcrysgro.2016.11.120} {\bibfield  {journal} {\bibinfo  {journal}
  {J. Cryst. Growth}\ }\textbf {\bibinfo {volume} {474}},\ \bibinfo {pages}
  {140} (\bibinfo {year} {2017})}\BibitemShut {NoStop}%
\bibitem [{\citenamefont {Shibuta}\ \emph {et~al.}(2015)\citenamefont
  {Shibuta}, \citenamefont {Oguchi}, \citenamefont {Takaki},\ and\
  \citenamefont {M.}}]{Shibuta-SR-2015}%
  \BibitemOpen
  \bibfield  {author} {\bibinfo {author} {\bibfnamefont {Y.}~\bibnamefont
  {Shibuta}}, \bibinfo {author} {\bibfnamefont {K.}~\bibnamefont {Oguchi}},
  \bibinfo {author} {\bibfnamefont {T.}~\bibnamefont {Takaki}}, \ and\ \bibinfo
  {author} {\bibfnamefont {O.}~\bibnamefont {M.}},\ }\href {\doibase
  10.1038/srep13534} {\bibfield  {journal} {\bibinfo  {journal} {Sci. Rep.}\
  }\textbf {\bibinfo {volume} {5}},\ \bibinfo {pages} {13534} (\bibinfo {year}
  {2015})}\BibitemShut {NoStop}%
\end{thebibliography}%

%\appendix
%\clearpage
%\onecolumngrid
%\renewcommand\thefigure{S\arabic{figure}}    
%\setcounter{figure}{0}
%\renewcommand\thetable{S\arabic{table}}    

\end{document}